\documentclass[conference]{./pvsctran}
\input epsf
\usepackage{graphicx}
\usepackage{hyperref}
\usepackage{graphicx}
\usepackage{amssymb}
\usepackage{amsmath}
\usepackage[super]{nth}
\usepackage{multicol}

\usepackage{cite}

\ifCLASSINFOpdf
\else
\fi
\begin{document}
%
\title{\LARGE Minority-carrier dynamics in semiconductors probed by two-photon
microscopy}

\author{\IEEEauthorblockN{Benoit Gaury}
\IEEEauthorblockA{Center for Nanoscale Science and Technology \\ National
Institute of Standards and Technology\\ Gaithersburg, MD 20899, USA}
\IEEEauthorblockA{Maryland NanoCenter, University of Maryland \\College Park, MD
20742, USA}
\and
\IEEEauthorblockN{Paul M. Haney}
\IEEEauthorblockA{Center for Nanoscale Science and Technology \\ National
Institute of Standards and Technology\\ Gaithersburg, MD, 20899, USA}
}


%


\def\be{\begin{equation}}
\def\ee{\end{equation}}

\setlength{\columnsep}{0.25in}

\maketitle

\begin{abstract}
Two-photon time-resolved photoluminescence has been recently applied to various
semiconductor devices to determine carrier lifetime and surface recombination
velocities. So far the theoretical modeling activity has been mainly limited to
the commonly used one-photon counterpart of the technique. Here we provide the
analytical solution to a 3D diffusion equation that describes two-photon
microscopy in the low-injection regime. We focus on a system with a single
buried interface with enhanced recombination, and analyze how transport, bulk
and surface recombinations influence photoluminescence decays. We find that bulk
measurements are dominated by diffusion at short times and by bulk recombination
at long times. Surface recombination modifies bulk signals when the optical spot
is less than a diffusion length away from the probed interface. In addition, the resolution
is increased as the spot size is reduced, which however makes the signal more
sensitive to diffusion.
\end{abstract}

%
\IEEEpeerreviewmaketitle

\section{Introduction}

The development of semiconductor devices such as photovoltaic solar cells
requires quantitative characterization of materials parameters to improve their
overall performances. While the minority-carrier lifetime may be the most
influential parameter for photovoltaic devices, polycrystalline materials such
as CdTe have many grains and grain boundaries whose contributions to
recombination remains unclear. Time-resolved photoluminescence (TRPL) is a
long-standing optical technique capable of probing bulk lifetime and surface
recombination velocities of direct bandgap materials. A TRPL experiment consists
of generating electron-hole pairs via a laser pulse, and collecting the
radiatively emitted photons over time. The non-radiative recombination occurring
in the bulk and at various interfaces influences the resulting photoluminescence
(PL) intensity
decay, so that adequate modeling of this decay can lead to values of bulk
lifetime, diffusion constant and surface recombination velocities.

Two different setups have been used to realize TRPL experiments. The first, most
common one, uses incident photons with energy larger than the semiconductor
bandgap. The absorption of a single photon is sufficient to generate an
electron-hole pair which leads to carrier generation that decays exponentially
away from the sample surface. A second route taken
in~\cite{Wang_1998}-\cite{Ma_2013} uses photons with energy smaller than the
semiconductor bandgap, that is in a spectral range where the material is normally transparent. The generation of an electron-hole pair now requires the
absorption of two photons. Because this non-linear process is proportional to 
the square of the incoming photon flux, the generation of carriers occurs
preferably at the focal volume of the optical setup. By changing the position of
the sample with respect to this focal volume, electron-hole
pairs can be generated far below the sample surface. Details on the operating
principle can be found in the
literature~\cite{Barnard_2013}-\cite{Kuciauskas2014}. There exists an
extensive literature on the modeling of one-photon TRPL~\cite{Boulou_1977,
Hooft_1986, Ahrenkiel_1989}, which assumes a carrier generation that decays
exponentially away from the sample surface. The direct carrier generation
below the surface taking place in a two-photon TRPL experiment requires new investigations.
\begin{figure}
    \begin{center}
    \includegraphics[width=0.23\textwidth]{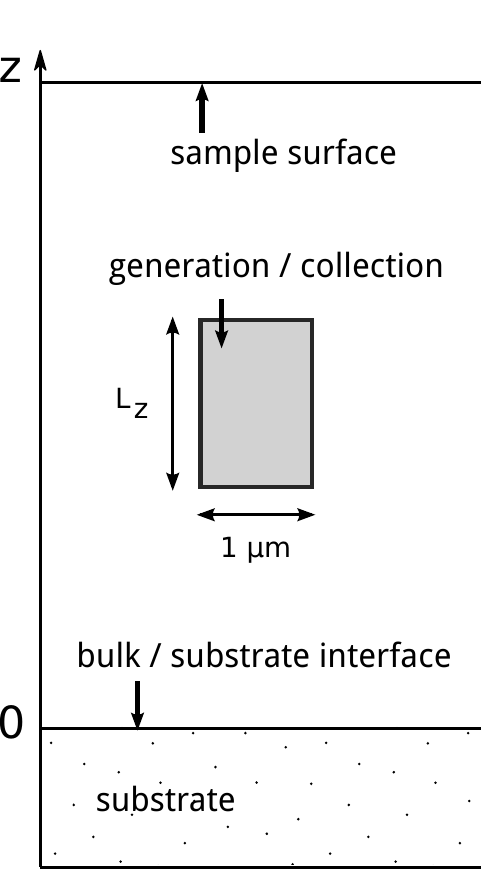}
    \end{center}
    \caption{\label{system} Schematic of the sample probed by two-photon
    microscopy. The optical spot (gray area) has lateral size $1~\rm \mu m$ and
    axial size $L_z=3~\rm \mu m$. Generation and collection regions are taken
    identical.} 
\end{figure}

In this manuscript we review our 3D model for two-photon TRPL
(section~\ref{model}), and we apply it
to optical spots in the bulk and at the buried bulk/substrate interface of our model system
described in Fig.~\ref{system}. 
In section~\ref{sec:diffusion} we focus on the
impact of carrier diffusion and recombination on PL decays for a bulk
measurement. We show how surface recombination changes PL intensities and
discuss the resolution of the optical technique in section~\ref{sec:surface}.
Throughout this paper, unless explicitly stated otherwise, we consider an
optical spot shaped as a rectangular prism with lateral dimensions of $1~\rm
\mu m$ and axial size $L_z=3~\rm \mu m$. We assume a uniform excitation of
carriers and identical generation and collection regions. While these two
regions may be in general different, this case can be obtained in an experimental setup
using a confocal microscope.

\section{Optically induced minority-carrier transport}
\label{model}
We start with our model for the transport of optically induced carriers in a
p-type material, and introduce the general solution of the problem for the case
of a single surface with enhanced recombination. This surface can describe a
sample surface or a buried interface as described by our model
system in Fig.~\ref{system}. We refer to~\cite{Gaury_2016} and references therein for more details on the derivations.

\bigskip
The minority-carrier transport is described by the 3D time-dependent diffusion equation
\be
    \frac{\partial n}{\partial t}(\mathbf{r},t)- D\Delta n(\mathbf{r},t) +
    \frac{n(\mathbf{r},t)}{\tau} = g(\mathbf{r})\delta(t),
    \label{3Dproblem}
\ee
and the boundary conditions determined by the surface recombination velocity $S$
\begin{align}
    \label{BC1}
    &D\frac{\partial n}{\partial z} = S n, \ \ z=0,\\
    &n = 0, \ \ z\rightarrow+\infty,
    \label{BC2}
\end{align}
where $n(\mathbf{r}, t)$ is the electron density, $D$ is the electron diffusion
constant, $\tau$ the bulk lifetime and $g(\mathbf{r})\delta(t)$ is the carrier
generation, taken to be instantaneous. Because we linearized the recombination rate
[third term in Eq.~(\ref{3Dproblem})],
Eq.~(\ref{3Dproblem}) is valid only for small excited carrier densities
(low-injection regime). We introduce the Green's function of the problem
$G(\mathbf{r}, \mathbf{r'}, t)$ that satisfies
\be
    \frac{\partial G}{\partial t}(\mathbf{r}, \mathbf{r'},t)- D\Delta G(\mathbf{r}, \mathbf{r'},t) +
    \frac{G(\mathbf{r}, \mathbf{r'},t)}{\tau} = \delta(\mathbf{r}-\mathbf{r'})
    \delta(t).
\label{GF}
\ee
Upon solving Eq.~(\ref{GF}) with the boundary conditions Eqs.~(\ref{BC1}) and
(\ref{BC2}), one finds~\cite{Gaury_2016}
\begin{align}
    G(x,&x',y,y',z,z',t) = \frac{e^{-t/\tau}}{2}  \frac{e^{-\frac{(x-x')^2}{4
    Dt}}}{2\sqrt{\pi Dt}} \frac{e^{-\frac{(y-y')^2}{4
    Dt}}}{2\sqrt{\pi Dt}} \nonumber\\
    &\times\Bigg[ \frac{e^{-\frac{(z-z')^2}{4Dt}} +
    e^{-\frac{(z+z')^2}{4Dt}}}{\sqrt{\pi Dt}} \nonumber\\
    & - 2\frac{S}{D}
    e^{{\frac{S}{D}(z+z')}+\frac{S^2}{D}t} \mathrm{erfc}\left(\frac{z+z'}{2\sqrt{Dt}} +
    S\sqrt{\frac{t}{D}}  \right) \Bigg],
    \label{G_final}
\end{align}
where $\mathrm{erfc}$ is the complementary error function. The electron density
follows by computing the convolution of the Green's function above with the
carrier generation profile. Because the optical
generation is assumed uniform, the electron density is obtained by simply integrating
the previous Green's function over the volume of the spot $V_{\rm spot}$
\be
    n(\mathbf{r}, t) = \int_{V_{\rm spot}} \mathrm{d}\mathbf{r'} \ G(\mathbf{r},
    \mathbf{r'}, t),
\label{convol}
\ee
and similarly integrating the above density over the collection volume $V_{\rm spot}$ yields the PL intensity
\be
    I(t) \propto \int_{V_{\rm spot}} \mathrm{d}\mathbf{r} \ n(\mathbf{r},t).
    \label{PL_signal}
\ee
The calculation of PL decays hence requires the integration of the Green's function
Eq.~(\ref{G_final}) over all spatial arguments over the generation/collection
volumes. We performed these integrations numerically for all the results
presented in this paper.

\bigskip
In addition to being limited to the low-injection regime, our model does not
include photon recycling~\cite{Stern_1974} and space charge effects caused by local
electric fields. Effects of differences in electron and hole mobilities as well
as the high injection regime have been studied numerically~\cite{Kanevce_2015}.

\section{Transport and lifetime effects in bulk measurements}
\label{sec:diffusion}
We first consider the case of a bulk measurement, i.e. the generation/collection
volumes are far from both the sample surface and the bulk/substrate interface.
We analyze how the carrier recombination and diffusion away from the
generation/collection region affect the PL intensity. 

\begin{figure}[b]
    \includegraphics[width=0.48\textwidth]{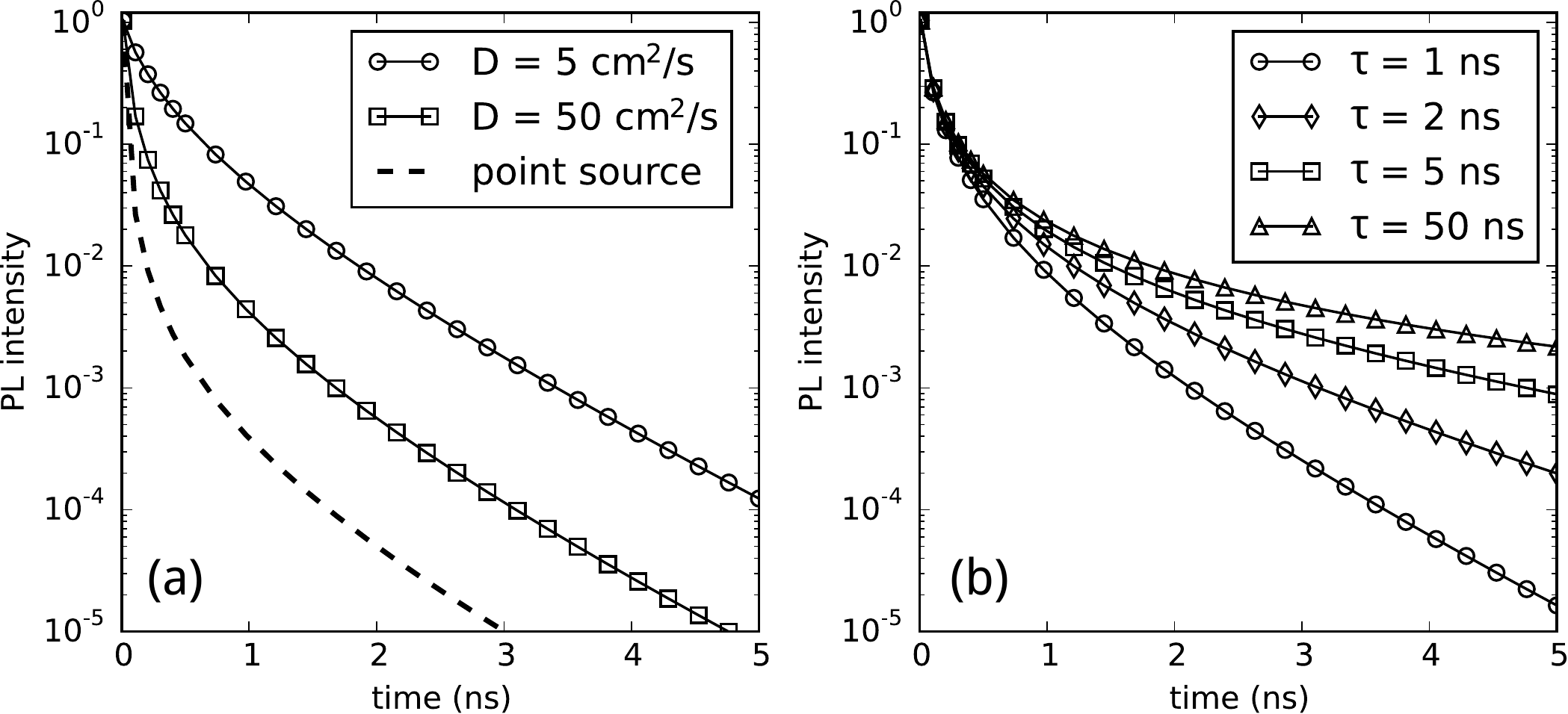}
    \caption{\label{diffusion} Normalized PL intensities as a function of time
    for the optical spot in the bulk. (a) The diffusion constant $D$ is varied:
    $D=5~\rm cm^2/s$ (circles), $D=50~\rm cm^2/s$ (squares)
    with $\tau=1~\rm ns$.  The dashed line
    corresponds to Eq.~(\ref{point_source}). (b) The lifetime is varied:
    $\tau=1~\rm ns$ (circles), $\tau=2~\rm ns$ (diamonds), $\tau=5~\rm ns$
    (squares), $\tau=50~\rm ns$ (triangles),  with $D=25~\rm cm^2/s$. } 
\end{figure}

\bigskip
Fig.~\ref{diffusion} shows the PL intensities obtained for (a) several values of the
diffusion constant, keeping $\tau=1~\rm ns$, and (b) several values of the bulk
lifetime, keeping $D=25~\rm cm^2/s$. We compare these results to the
limiting case of a point source whose PL decay reads~\cite{Gaury_2016}
\be
I_b(t) \propto \frac{e^{-t/\tau}}{8(\pi Dt)^{3/2}}.
\label{point_source}
\ee
While the PL intensities at short times ($t<\tau$) are all identical in
Fig.~\ref{diffusion}(b), sharp drops occur as the diffusion constant, hence the
diffusion velocity, is
increased in Fig.~\ref{diffusion}(a). These drops reveal that charge carriers
diffuse away from the generation spot and recombine outside of the collection
volume, so that emitted photons are not collected. Note that the fast decay is
not exponential but algebraic, as shown by Eq.~(\ref{point_source})
($e^{-t/\tau}\approx 1$). One can recover purely exponential decays at short
times when collecting all emitted photons, which implies a collection region
much larger than the diffusion length. Differences in generation and collection
volumes can therefore be used as a knob to characterize carrier diffusion
properties.

At long times ($t>\tau$), Fig.~\ref{diffusion}(a) shows identical exponential
decays, while the slopes of the decays decrease as the bulk lifetime is
increased in Fig.~\ref{diffusion}(b). Eq.~(\ref{point_source}) shows that
recombination exponentially reduces the PL intensity. At long times, this
exponential decay (recombination) dominates the previously discussed algebraic
decay (diffusion). This is seen
in Fig.~\ref{diffusion}(b) for $\tau=1~\rm ns$ and $\tau=2~\rm ns$, while the
traces for $\tau=5~\rm ns$ and $\tau=50~\rm ns$ are still in the diffusion
dominated regime. As a result, experimentally, in the long time limit a mono-exponential fit
should suffice to determine the bulk lifetime $\tau$. Comparisons with
experimental data can be found in~\cite{Gaury_2016}.

\section{Surface recombination effects and resolution of 2-photon TRPL}
\label{sec:surface}
We turn to calculations done for a generation/collection region at or close to
the bulk/substrate interface. We analyze how the enhanced recombination at the
interface changes the previous bulk PL decays, and discuss the resolution of the
two-photon TRPL technique.  Plots presented in this section were obtained with
$\tau=1~\rm ns$ and $D=25~\rm cm^2/s$. We denote $z$ as the distance of the bottom of
the optical spot from the bulk/substrate interface.

\bigskip
\begin{figure}[b]
    \includegraphics[width=0.48\textwidth]{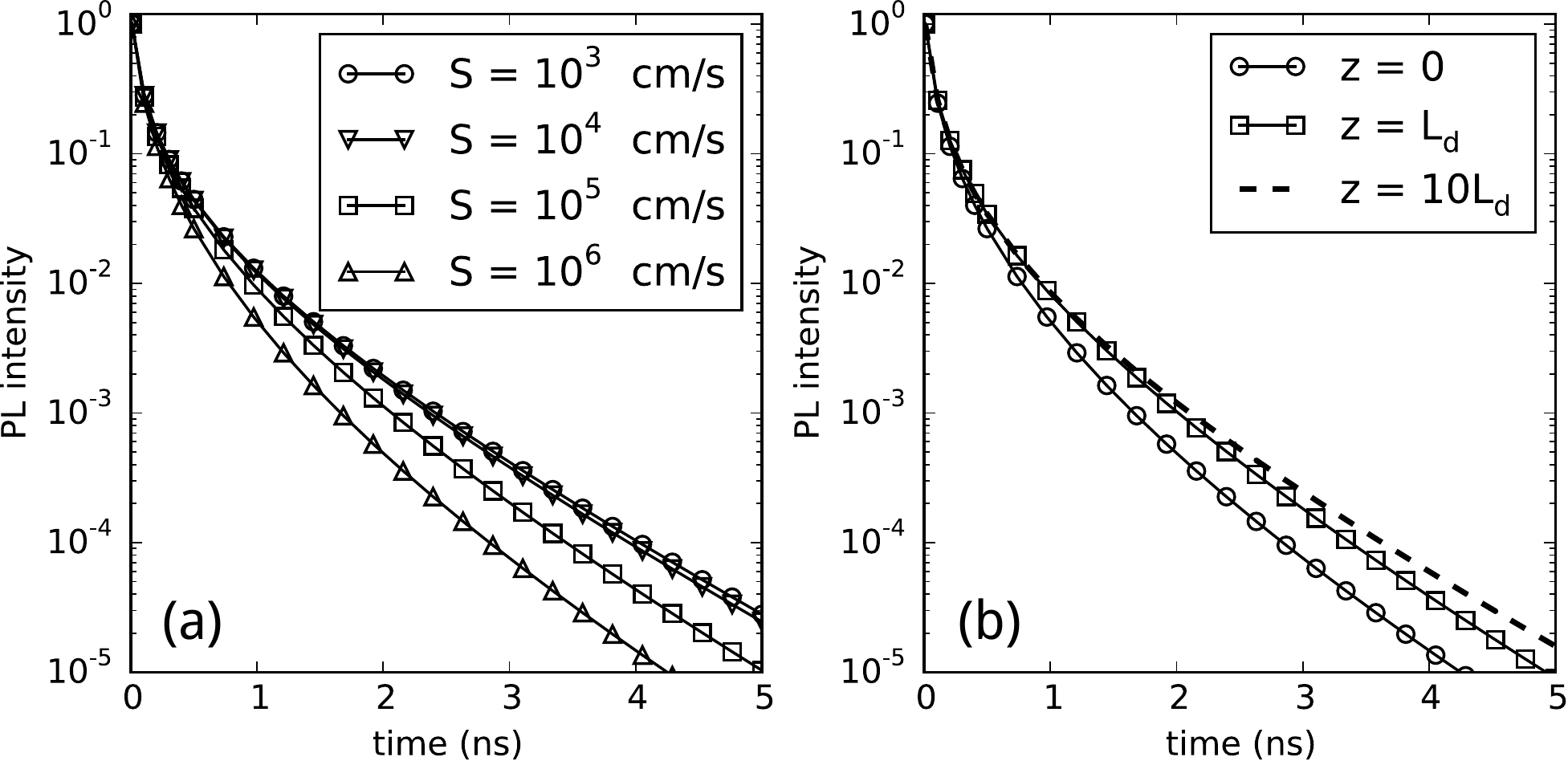}
    \caption{\label{surface} Normalized PL intensities as a function of time
    for the optical spot around the bulk/substrate interface. (a) The spot is at the
    bulk/substrate interface ($z=0$) and the surface recombination
    velocity varies: $S=10^3~\rm cm/s$ (circles), $S=10^4~\rm cm/s$ (down triangles),
    $S=10^5~\rm cm/s$ (squares), $S=10^6~\rm cm/s$ (up triangles).  (b) The
    distance of the bottom of the optical spot is changed: $z=0$ (circles),
    $z=L_d$ (squares), $z=10L_d$ (dashed line). Calculations were done with
    $S=10^6~\rm cm/s$. For all plots $\tau=1~\rm ns$ and $D=25~\rm cm^2/s$.} 
\end{figure}
Fig.~\ref{surface}(a) shows PL intensities for various surface recombination
velocities when the generation/collection spot is at the bulk/substrate
interface ($z=0$). As expected the PL signal decreases more rapidly for greater
values of $S$, as the recombination current is increased at the interface.  PL
decays at short times are still dominated by diffusion, as the fast decays for
$t<1~\rm ns$ are independent of the value of the recombination velocity.
However, contrary to the previous bulk calculations, PL decays are not purely exponential
for times $t>\tau$ as $S$ is increased. Consequently, a numerical fitting
procedure relying on Eq.~(\ref{G_final}) together with known generation profile and collection
volume is necessary to determine materials parameters.

We increased the distance of the
generation/collection region from the bulk/substrate interface in
Fig.~\ref{surface}(b). Calculations were done with $z=0$ (bottom of the spot at
the interface), $z=L_d$ ($L_d=\sqrt{D\tau}$: diffusion length) and $z=10L_d$.
The latter can be considered in the bulk of the system (i.e. no surface effects). We
find that a displacement of the optical spot away from the interface by a diffusion
length ($L_d=1.6~\rm \mu m$) is enough to make the PL intensity insensitive to
the probed interface. 

\begin{figure}
    \includegraphics[width=0.48\textwidth]{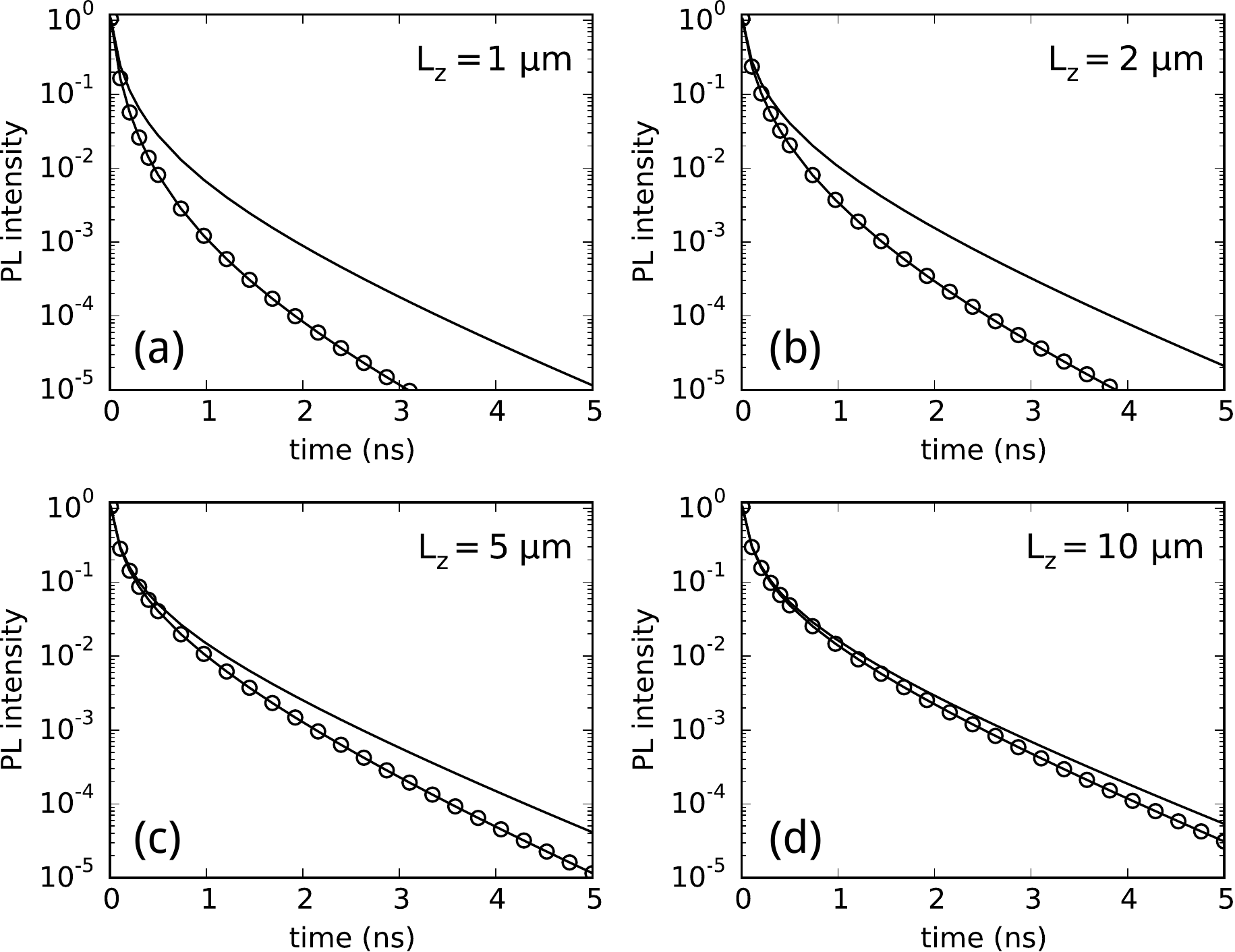}
    \caption{\label{resolution} Normalized PL intensities as a function of time
    when the axial size of the spot ($L_z$) varies. (a) $L_z=1~\rm \mu m$, (b)
    $L_z=2~\rm \mu m$, (c) $L_z=5~\rm \mu m$, (d) $L_z=10~\rm \mu m$. 
    The lines with symbols correspond to $S=5\times 10^5~\rm cm/s$ and the
    simple lines to $S=0$. All
    calculations were done with $\tau=1~\rm ns$ and $D=25~\rm cm^2/s$.} 
\end{figure}

\bigskip
In Fig.~\ref{resolution}, we examine the variation in the PL decays as a function
of the axial size of the generation/collection
region. $L_z$ is varied from $1~\rm \mu m$ to $10~\rm \mu m$,
and we compare the PL decays to the case $S=0$ (lines without symbols). The spot is
at the bulk/substrate interface ($z=0$).
Increasing the spot size should reduce the effect of diffusion. However, because
we only increase the axial dimension while keeping the lateral size ($1~\rm \mu
m$) smaller than the diffusion length ($1.6~\rm \mu m$), there is still a rapid
drop of the PL decay for $t<1~\rm ns$. The drop is stronger for $L_z=1~\rm \mu
m$, and the PL intensity is extremely reduced when the surface effects become
visible, which may pose a challenge for experimental detection of photons. At times $t>\tau$ comparisons to the
case $S=0$ show that the PL signals converge towards a single exponential decay
as $L_z$ is increased. As expected the contribution of the bulk to recombination
becomes dominant as the surface to volume ratio of the spot decreases.

\section{Conclusion}
Two-photon microscopy is a non-invasive and non-destructive optical technique
that can probe subsurface materials parameters. Our 3D modeling allowed us to
separate lifetime (recombination) from transport (diffusion) effects in PL
intensities. In particular we found that the former influences decays at long
times while the latter dominates short times. We have shown that the resolution
of the technique for interfacial features is improved as the optical spot is confined at the bulk/substrate
interface. However this comes with an increase of the diffusion effects causing
a sharp drop in the PL decay at short times.


\section*{Acknowledgment}
B.~Gaury acknowledges support under the Cooperative Research
Agreement between the University of Maryland and the National Institute of
Standards and Technology Center for Nanoscale Science and Technology, Award
70NANB10H193, through the University of Maryland.

\vfil\eject

\bibliographystyle{IEEEtran}
\bibliography{IEEEabrv,references}

\begin{thebibliography}{10}
\providecommand{\url}[1]{#1}
\csname url@samestyle\endcsname
\providecommand{\newblock}{\relax}
\providecommand{\bibinfo}[2]{#2}
\providecommand{\BIBentrySTDinterwordspacing}{\spaceskip=0pt\relax}
\providecommand{\BIBentryALTinterwordstretchfactor}{4}
\providecommand{\BIBentryALTinterwordspacing}{\spaceskip=\fontdimen2\font plus
\BIBentryALTinterwordstretchfactor\fontdimen3\font minus
  \fontdimen4\font\relax}
\providecommand{\BIBforeignlanguage}[2]{{%
\expandafter\ifx\csname l@#1\endcsname\relax
\typeout{** WARNING: IEEEtran.bst: No hyphenation pattern has been}%
\typeout{** loaded for the language `#1'. Using the pattern for}%
\typeout{** the default language instead.}%
\else
\language=\csname l@#1\endcsname
\fi
#2}}
\providecommand{\BIBdecl}{\relax}
\BIBdecl

\bibitem{Wang_1998}
H.~Wang, K.~S. Wong, B.~A. Foreman, Z.~Y. Yang, and G.~K.~L. Wong, ``One- and
  two-photon-excited time-resolved photoluminescence investigations of bulk and
  surface recombination dynamics in {ZnSe},'' \emph{J. Appl. Phys.}, vol.~83,
  no.~9, pp. 4773--4776, 1998.

\bibitem{Ma_2013}
J.~Ma, D.~Kuciauskas, D.~Albin, R.~Bhattacharya, M.~Reese, T.~Barnes, J.~V. Li,
  T.~Gessert, and S.-H. Wei, ``Dependence of the minority-carrier lifetime on
  the stoichiometry of {CdTe} using time-resolved photoluminescence and
  first-principles calculations,'' \emph{Phys. Rev. Lett.}, vol. 111, p.
  067402, 2013.

\bibitem{Barnard_2013}
E.~S. Barnard, E.~T. Hoke, S.~T. Connor, J.~R. Groves, T.~Kuykendall, Z.~Yan,
  E.~C. Samulon, E.~D. Bourret-Courchesne, S.~Aloni, P.~J. Schuck, C.~H.
  Peters, and B.~E. Hardin, ``Probing carrier lifetimes in photovoltaic
  materials using subsurface two-photon microscopy,'' \emph{Sci. Rep.}, vol.~3,
  pp. 2098--2106, 2013.

\bibitem{Kuciauskas2014}
D.~Kuciauskas, S.~Farrell, P.~Dippo, J.~Moseley, H.~Moutinho, J.~V. Li,
  A.~M.~A. Motz, A.~Kanevce, K.~Zaunbrecher, T.~A. Gessert, D.~H. Levi, W.~K.
  Metzger, E.~Colegrove, and S.~Sivananthan, ``Charge-carrier transport and
  recombination in heteroepitaxial {CdTe},'' \emph{J. Appl. Phys.}, vol. 116,
  pp. 123\,108--123\,115, 2014.

\bibitem{Boulou_1977}
M.~Boulou and D.~Bois, ``Cathodoluminescence measurements of the
  minority--carrier lifetime in semiconductors,'' \emph{J. Appl. Phys.},
  vol.~48, no.~11, pp. 4713--4721, 1977.

\bibitem{Hooft_1986}
G.~W. ‘t Hooft and C.~van Opdorp, ``Determination of bulk minority‐carrier
  lifetime and surface/interface recombination velocity from photoluminescence
  decay of a semi‐infinite semiconductor slab,'' \emph{J. Appl. Phys.},
  vol.~60, no.~3, pp. 1065--1070, 1986.

\bibitem{Ahrenkiel_1989}
R.~K. Ahrenkiel and D.~J. Dunlavy, ``Minority-carrier lifetime in
  {${\mathrm{Al}}_{\mathit{x}}$${\mathrm{Ga}}_{1\mathrm{-}\mathit{x}}$As},''
  \emph{J. Vac. Sci. Tech. A}, vol.~7, no.~3, pp. 822--826, 1989.

\bibitem{Gaury_2016}
B.~Gaury and P.~M. Haney, ``Probing surface recombination velocities in
  semiconductors using two-photon microscopy,'' \emph{J. Appl. Phys.}, vol.
  119, pp. 125\,105--125\,113, 2016.

\bibitem{Stern_1974}
F.~Stern and J.~M. Woodall, ``Photon recycling in semiconductor lasers,''
  \emph{J. Appl. Phys.}, vol.~45, no.~9, pp. 3904--3906, 1974.

\bibitem{Kanevce_2015}
A.~Kanevce, D.~Kuciauskas, D.~H. Levi, A.~M. Allende~Motz, and S.~W. Johnston,
  ``Two dimensional numerical simulations of carrier dynamics during
  time-resolved photoluminescence decays in two-photon microscopy measurements
  in semiconductors,'' \emph{J. Appl. Phys.}, vol. 118, no.~4, pp.
  045\,709--045\,715, 2015.

\end{thebibliography}

\end{document}